\documentclass{npw}
\usepackage{graphicx,color}
\usepackage{colordvi}
\def\be{\begin{equation}}
\def\ee{\end{equation}}
\def\bea{\begin{eqnarray}}
\def\eea{\end{eqnarray}}
\def\l{\label}
\def\r{{\bf r}}
\def\p{{\bf p}}
\def\s{{\bf s}}
\def\om{\omega}

\def\sinh{\hbox{sinh}}

\def\exp{\hbox{exp}}

\def\sinh{\hbox{sinh}}
\def\Im{{\mbox {\rm Im}}}
\def\Re{{\mbox {\rm Re}}}

\def\eps{\varepsilon}
\def\siml{\hbox{\kern.1em \lower.6ex \hbox{$\sim$} \kern-1.12em
 \raise.6ex \hbox{$<$} \kern.1em}}
\def\simg{\hbox{\kern.1em \lower.6ex \hbox{$\sim$} \kern-1.12em
 \raise.6ex \hbox{$>$} \kern.1em}}

\begin{document}
\title{Semiclassical shell-structure moment of 
inertia within the phase-space approach}
\author{
                                                                \\[ 1.0ex]
D V Gorpinchenko and A G Magner\email{magner@kinr.kiev.ua}\\
{\it Institute for Nuclear Research, 03680 Kyiv, Ukraine}\\
{\it and National Technical Institute University of Ukraine, 03056, Kyiv} 
                                                                     \\[ 1.0ex]
J Bartel\\
{\it Universit\'e de Strasbourg, IPHC, 67037 Strasbourg, France}
                                                                     \\[ 1.0ex]
J P Blocki\\
{\it National Centre for Nuclear Research,  PL-00681 
Warsaw, Poland}
}

\date{}
\maketitle

\vspace*{-1.05cm}
\begin{abstract}

The moment of inertia for nuclear 
collective rotations was derived  
within the semiclassical  
approach based on the cranking model and the
Strutinsky shell-correction method by using
the non-perturbative periodic-orbit theory in 
the phase space variables. 
This moment of inertia for adiabatic 
(statistical-equilibrium) rotations can be approximated by
the generalized rigid-body moment of inertia 
accounting for the shell corrections of the
particle density. A semiclassical phase-space trace formula 
allows to express quite accurately 
the shell components of the moment of inertia  
in terms of the free-energy shell corrections for integrable and 
partially chaotic Fermi systems, 
in good agreement with the quantum calculations.
                                                                     \\[ 2.0ex]
\end{abstract}

\section{Introduction}

Many significant phenomena deduced from experimental data on nuclear 
rotations were explained
within theoretical approaches based mainly on the cranking model 
\cite{inglis,bohrmot,valatin,bohrmotbook} and the Strutinsky 
shell-correction method \cite{strutMOP,funnyhills}
extended by Pashkevich and Frauendorf \cite{pashfrau,mikhailovepnp}.  
For a deeper understanding of the correspondence
between classical and quantum physics of the shell components
of the moment of inertia (MI), it is worth to analyze 
them within the semiclassical periodic-orbit theory (POT) 
\cite{gutzpr,gutz,strut,strutmag,strutmagofdos1977,MKS1978,book}. 
See also Refs.\ \cite{KMS1979,MKS1979} for the semiclassical 
description of the so called 
``classical rotation'' as an alignment of the particle angular momenta 
along the symmetry axis, but also 
the magnetic susceptibilities in 
metallic clusters  and quantum dots \cite{FKMS1998}. 
The semiclassical perturbation expansion of Creagh  
\cite{book} was used in the POT calculations \cite{DFP2004} 
of the MI shell corrections for a spheroidal cavity mean field.
The non-perturbative Gutzwiller
 POT extended to the bifurcation phenomena 
\cite{MFAMMSB1999,MAFM2002,MAF2006,MYAB2011} was applied in \cite{MSKB2010}
within the cranking model for the 
adiabatic collective rotations (around an
axis perpendicular to the symmetry axis)  in the case of the
harmonic-oscillator mean field.
The MI $\Theta $ for the collective adiabatic 
(statistical-equilibrium) rotations was described as a sum of the 
Extended Thomas-Fermi (ETF) MI $\Theta _{\rm ETF} $ \cite{bartelETF1994} and 
shell corrections $\delta \Theta $ 
 \cite{MSKB2010,belyaev}).

In the present work, we obtain a semiclassical 
phase-space trace formula for the MI shell components $\delta \Theta $ in 
terms of the free-energy shell corrections $\delta F$ for integrable 
Hamiltonians, 
including those of the harmonic-oscillator mean field 
\cite{MSKB2010,magosc1978} and a 
spheroidal cavity \cite{MAFM2002,MYAB2011},
as well as for partially chaotic Fermi systems \cite{belyaev}. 

\section{Cranking model for nuclear rotations}

Within the cranking model, the nuclear collective rotation 
of the independent-particle Fermi system 
is associated with a 
many-body Hamiltonian (Routhian)     
\be\l{hamilcrm}
  H^{}_{\om} = H - \om \ell_x,
\ee
and its eigenvalue equation is given by
\be\l{eigencrm}
  H^{}_{\om} \, \psi_i^{\om} = \eps_i^{\om} \, \psi_i^{\om} \; .
\ee
The frequency (Lagrange multiplier) $\om$ is determined by the
angular momentum projection $I_x$ onto the $x$ axis (perpendicular
to the symmetry $z$ axis) through the constraint for $\om=\om(I_x)$,
\be\l{costraint}
\langle \ell_x \rangle^{}_{\om} \equiv
d_s \sum_i n_i^{\om} \int {\rm d} \r
\;\psi_i^{\om}\left(\r\right)\ell_x\;
\overline{\psi}_i^{\;\om}\left(\r\right)=I_x,
\ee
where $\ell_x$ and $n_i^{\om}$ are respectively the particle angular-momentum 
projection
onto 
the $x$ axis and the occupation numbers,
\be\l{occupnumb}
n_i^{\om}=\left[1+\exp\left(\frac{\eps_i^{\om}-\lambda^{\om}}{T}\right)\right]^{-1},
\ee
$T$ is the temperature. In the following 
$\lambda^{\om}$ is the chemical
potential, and  $d^{}_s$ the spin (spin-isospin) degeneracy. 
Brackets, like in $\langle \ell_x \rangle$
indicate a quantum average and a bar above the functions the complex
conjugation. 
For simplicity, the spin degree of freedom is included here 
only through the degeneracy factor $d^{}_s$, and then, 
the moment of inertia $\Theta_x$ can be considered as the following
susceptibility: 
\be\l{midef}
\Theta_{x}(\om)=
\frac{\partial \langle \ell_x\rangle_\om}{\partial \om}=
\frac{\partial^2 E(\om)}{\partial \om^2},
\ee
where
\be\l{energyom} 
E(\om)=\langle H\rangle_\om
\equiv d_s \sum_i n_i^{\om}\int {\rm d} \r
\;\psi_i^{\om}\; \left(\r\right) \;H \;
\overline{\psi}_i^{\;\om}\left(\r\right)
\ee
is the energy $E(I_x)$ of the yrast line due to the constraint 
(\ref{costraint}). For the particle number conservation, one has
\be\l{partconserv}
  N = d_s\sum_i n_i^{\om}
\int {\rm d} \r\; \psi_i^{\om}\;\overline{\psi}_i^{\om}=   
   d_s \sum_i n_i \; .    
                                                                    \\[ 1.0ex]
\ee
In the coordinate representation, Eq.\ (\ref{midef}) can be re-written for
the adiabatic rotations through the one-body Green's function $G$:
$$
\Theta_{x}=\frac{2 d_s}{\pi}\int^{\infty}_0 {\rm d}\eps\;
n({\eps}) \int {\rm d} \r_1 \int {\rm d} \r_2 \; 
\ell_{x}(\r_1)\; \ell_{x}(\r_2) \hspace{1.0cm}
$$
\be\l{micoorrep}
 \hspace{0.5cm} \times \, \Re \left[G\left(\r_1,\r_2;\eps \right) \right]\;
\Im \left[G\left(\r_1,\r_2;\eps\right)\right]\; .    
\ee
This representation is useful within the semiclassical POT, 
weakening the criterium 
of the quantum perturbation approximation: The rotation
excitation energy $\hbar \omega$ becomes smaller 
than the distance between major 
shells $\hbar \Omega$ 
($\hbar \Omega \!\approx\! \eps^{}_F/N^{1/3}$
with the Fermi energy $\eps^{}_F$). This is in contrast to the quantum criterium
of smallness of the excitation energies with respect to
the nearest neighbor single-particle (s.p.) level spacing around 
the Fermi surface.
                                                                    \\[ 10.0ex]
\begin{figure}
\begin{center}
\includegraphics[width=4.9cm,height=4.5cm,angle=0]{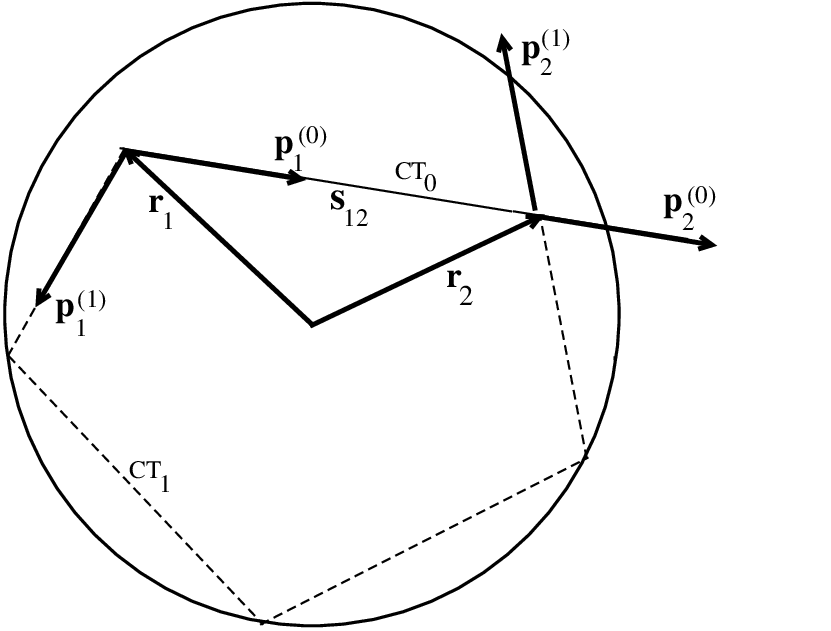}
\end{center}
%
\caption{
Classical trajectories (CT) from point $\r^{}_1$ with the
momentum $\p^{}_1$ to $\r^{}_2$, with $\p^{}_2$. Here 
$CT_0$ is the direct path, and $CT_1$ one with
reflection, and ${\bf s}^{}_{12}=\r^{}_1-\r^{}_2$.
}
\label{fig1}
\end{figure}
${}$
                                                                    \\[ -6.0ex]

\vspace{-1.4cm}
\section{Semiclassical Green's function and particle density}

For the s.p. 
Green's function $G\left(\r_{1},\r_{2};\eps\right)$, we shall use
the semiclassical Gutzwiller trajectory expansion \cite{gutz},
\bea\l{gfun}
 && G\left(\r^{}_{1},\r_{2};\eps\right) \approx
    G_{\rm scl}\left(\r_{1},\r_{2};\eps\right) = \sum_{\rm CT} G^{}_{\rm CT}, 
\nonumber\\
&& G^{}_{\rm CT} = {\cal A}_{\rm CT}\;
  \exp\left[\frac{i}{\hbar}S_{\rm CT}\left(\r_{1},\r_{2};\eps\right)
   - \frac{i \pi}{2}\mu^{}_{\rm CT}\right],
\eea
%
where the sum runs over the classical trajectories (CT) from 
point $\r^{}_1$
to point 
$\r^{}_2$ with the particle energy $\eps$ (see Fig.\ \ref{fig1}). 
${\cal A}_{\rm CT}\left(\r_{1},\r_{2};\eps\right)$ is 
the amplitude depending on the
CTs degeneracy (in the case $\r_1=\r_2$) 
and their stability propagator, 
$S_{CT}\left(\r_{1},\r_{2};\eps\right)$ is the particle action,
$\mu^{}_{\rm CT}$ the Maslov phase determined by 
the caustic and turning catastrophe
points along the CT \cite{gutz,strut,strutmag,book,MAF2006}.

There are several reasons leading to oscillations of the MI in Eq. 
(\ref{micoorrep}), local and non-local, where only the local part is related 
to the 
shell effects, whereas non-local contributions could have their origin in 
reflections of the particle from the boundary. 
The nearly local approximation ($|\s^{}_{12}| = | \r^{}_2 \!-\! \r^{}_1 | 
\ll R$) is valid after a statistical averaging over many microscopic 
quantum states. 
Then, the small parameter, which is the product of the two dimensionless
quantities $\;\om/\Omega$ and $\;S/\hbar$, used in \cite{DFP2004} for 
applying the 
perturbation approach of Creagh to the classical dynamics \cite{book} 
becomes  in our approach more weak,  
$(\om/\Omega)\;S/\hbar \simg 1$, under the usual 
semiclassical condition  $S/\hbar \sim k^{}_F R\sim N^{1/3} \gg 1$, 
where $k^{}_F=p^{}_F/\hbar$ is the Fermi momentum in $\hbar$ units and 
$R$ is the mean nuclear radius. According to (\ref{gfun}), one  
can split the Green's function $G(\r^{}_{1},\r_{2};\eps)$
into a contribution $G^{}_{{\rm CT}_0}$ coming from the direct path between 
the two points and a contribution $G^{}_{{\rm CT}_1}$ that contains the 
contributions from all other trajectories involving reflections
\be\l{Gsplit}
  G_{\rm scl} = G^{}_{{\rm CT}_0} + G^{}_{{\rm CT}_1},
\ee
where $G^{}_{{\rm CT}_0}$ is the component related to the trajectory
CT$_0$ for which the action $S^{}_{{\rm CT}_0}$ 
disappears in the limit $\r_2 \to \r_1$. One obtains 
\be\l{G0} 
 G^{}_{{\rm CT}_0} = - \frac{m}{2 \pi \hbar^2 s^{}_{12}} 
                \exp\left[\frac{i}{\hbar} s^{}_{12} p(\r) \right],
\ee
with the nucleon mass $m$, the particle momentum 
$\,\, p(\r) = \sqrt{2m[\eps - V(\r)]\,}$, $\;\r=(\r_1+\r_2)/2$,
and the potential $V(\r)$.
                                                                   \\[ -1.0ex]
 
For a semiclassical 
statistical equilibrium rotation with  
constant frequency $\om$,
one finds \cite{MSKB2010,belyaev} according to (\ref{micoorrep}), 
\be\l{misplit}
\Theta \approx \Theta_{\rm GRB} = 
m \int {\rm d} \r \; r_\perp^2 \, \rho_{\rm scl}(\r) 
= \Theta_{\rm ETF} + \delta \Theta_{\rm scl},  
\ee
where $ \Theta_{\rm GRB}=\Theta_{\rm GRB}^{\rm ETF} + 
\delta \Theta_{\rm GRB}^{\rm scl}$ is 
the generalized rigid-body (GRB) MI,
with \cite{bartelETF1994,belyaev}
\be\l{dthetaETF}
\Theta_{\rm ETF} \approx \Theta_{\rm GRB}^{\rm ETF}=
m \int {\rm d}\r\; r_\perp^2 \, \rho^{}_{ETF}\left(\r\right)
\ee
and $ \delta \Theta_{\rm scl} $ 
its shell correction \cite{MSKB2010,belyaev},
\be\l{dtsclsplit}
\delta \Theta_{\rm scl} \approx \delta \Theta_{\rm GRB}^{\rm scl}=
m \int {\rm d}\r \; r_\perp^2 \, \delta \rho_{\rm scl}(\r)
\ee
with $ r_\perp^2 = y^2 + z^2 $. 
Such a splitting (\ref{misplit}) is associated with 
that of the spatial particle density
$\rho(\r)$, 
\be\l{partdensplit} 
\rho(\r) = 
 - \frac{1}{\pi}\;\Im \int {\rm d} \eps\; n\left(\eps\right)
\left[G\left(\r_1,\r_2;\eps\right)\right]_{\r_1=\r_2=\r},
\ee
and the one of equation 
(\ref{Gsplit}) in terms of its ETF particle density
$\rho^{}_{\rm ETF}$ and its shell
correction $\delta \rho_{\rm scl}(\r)$,
\be\l{rhoscl}
\rho^{}_{\rm scl}\left(\r\right)=
\rho^{}_{\rm ETF} + \delta \rho_{\rm scl}(\r),
\ee
where 
\be\l{rhotf}
\rho^{}_{\rm ETF}\left(\r\right) \!=\!
- \frac{1}{\pi}\,\Im \! \int \! {\rm d} \eps\; \widetilde{n}\left(\eps\right)
\left[G_{{\rm CT}_0}\left(\r_1,\r_2;\eps\right)\right]^{}_{\r_1=\r_2=\r}
\ee
and 
\be\l{drhoscl}
  {}\hspace{-0.2cm}
\delta \rho_{\rm scl}\left(\r\right) \!=\!
-\frac{1}{\pi} \,\Im \! \int \!\! {\rm d} \eps\; \delta n\left(\eps\right) \!
\left[G^{}_{{\rm CT}_1} \!\left(\r_1,\r_2;\eps\right)\right]^{}_{\r_1=\r_2=\r} .
\ee
The standard decomposition of the occupation numbers $n=\widetilde{n}+\delta n$
into the smooth and fluctuating parts is used as usually in the 
shell correction method \cite{funnyhills}.

\section{Phase space trace formulas}

Substituting (\ref{partdensplit}) and (\ref{gfun}) into the integrand 
(\ref{misplit}),  for the total semiclassical MI $\Theta_x$, 
one obtains the phase-space trace formula:
                                                                   \\[ -1.0ex]
$$
 \Theta_{\rm scl} \approx d_s \, m \!\int\! {\rm d} \eps\; \eps\; n(\eps)
\int \frac{{\rm d} \r_2 {\rm d} \p_1}{(2 \pi \hbar)^3}\;
\frac{r_{\perp}^2}{\eps} \, f_{\rm scl}(\r_2,\p_1,\eps) 
$$
\be\l{misclps}
= \Theta_{\rm ETF} +\delta \Theta_{\rm scl},
\ee
where 
\be\l{fscl}
f_{\rm scl}(\r_2,\p_1,\eps) = G_{\rm scl}(\r_2,\p_1,\eps) 
\exp\left[i \, \p_1\left(\r_1\!-\!\r_2\right)/\hbar\right]
\ee
and where now $r_{\perp}^2=y_2^2+z_2^2$. Here 
$G_{\rm scl}(\r_2,\p_1,\eps)$ is a 
semiclassical Green's function
in the mixed phase-space representation obtained by the
Fourier transformation of
the Green's function 
  $G_{\rm scl}(\r_1,\r_2,\eps)$, Eq.\ (\ref{gfun}),
\bea\l{psgfun} 
G_{\rm scl}(\r_2,\p_1,\eps){}\hspace{-0.3cm}
&=& {}\hspace{-0.3cm} \Re \sum_{\rm CT}
  \Big|J_{\rm CT}\left(\p_1^{\perp},\p_2^{\perp}\right)\Big|^{1/2}
  \delta\left[\eps-H(\r_2,\p_2)\right] \nonumber\\
&\times&
\exp\left\{\frac{i}{\hbar} 
S_{\rm CT}\left(\r_1,\r_2,\eps\right) \!-\!
\frac{i\pi}{2}\mu^{}_{\rm CT}\right\},
\eea
where $J_{\rm CT}\left(\p_1^{\perp},\p_2^{\perp}\right)$ is the Jacobian 
for the transformation from the perpendicular-to-CT momentum
$\p_1^\perp$ to the $\p_2^\perp$ one. 
We inserted formally the additional integral over $\r_1$ with the 
$\delta(\r_2-\r_1)$
function in (\ref{misplit}) and transformed the spacial coordinates 
$\r_1$ and $\r_2$ to the 
phase-space variables $\r_2$ and $\p_1$ \cite{MAF2006,belyaev}.  Using then the
Fourier transformation of this $\delta$ function from the coordinate difference
$\r_2-\r_1$ to a new momentum ${\tilde \p}$ and integrating
the MI in such a phase space representation over perpendicular-to-CT components
of ${\tilde \p}$ by the stationary phase 
method, one arrives at (\ref{misclps}).
Note that the ETF component of the Green's function CT$_0$ (\ref{Gsplit})
under the nearly local approximation ($\r_1\rightarrow \r_2 \rightarrow \r$ and 
$\p_1 \rightarrow \p_2 \rightarrow \r$)
is related to the Thomas-Fermi spectral function (\ref{fscl})
\be\l{distrfuntf}
f_{\rm scl}(\r,\p,\eps)  
\rightarrow f^{}_{\rm TF}(\r,\p,\eps) = \delta(\eps-H(\r,\p)).
\ee

Equation (\ref{misclps}) looks similar (the same besides of the factor 
$m r_{\perp}^2/\eps$) to the semiclassical s.p. energy $E_{\rm scl}$,
\bea\l{escl}
E_{\rm scl}&=& d_s\int {\rm d} \eps\; \eps\; n(\eps) g_{\rm scl}(\eps)
= d_s\int {\rm d} \eps\; \eps\;n(\eps) 
\int \frac{{\rm d} \r_2 {\rm d} \p_1}{(2 \pi \hbar)^3}
\nonumber\\
&\times& f_{\rm scl}(\r_2,\p_1,\eps)   \approx  E_{\rm ETF} + \delta E_{\rm scl},
\eea
where $E_{\rm ETF}$ is the ETF energy and $\delta E_{\rm scl}$ the energy 
shell correction. We used also
the phase-space trace formula for the semiclassical level density 
$g_{\rm scl}(\eps)$ \cite{gutz,strut,strutmag,book,MAF2006,MYAB2011}
with a similar decomposition,
\be\l{gscldef}
  g_{\rm scl}(\eps) = \int \frac{{\rm d} \r_2 {\rm d} \p_1}{(2 \pi \hbar)^3}\; 
      f_{\rm scl}(\r_2,\p_1,\eps) \hspace{2.0cm}
\ee
\be\l{gscl}
 \hspace{2.0cm}
\approx g^{}_{\rm ETF}(\eps) + \delta g_{\rm scl}(\eps) \; .
\ee
Therefore, multiplying and dividing identically (\ref{misclps})
by the energy $E_{\rm scl}$ (\ref{escl}), one finally arrives at
\be\l{MIE} 
\Theta_{\rm scl} \approx  
          m \, \langle \frac{r_{\perp} ^2}{\varepsilon} \rangle \, E_{\rm scl},
\ee
where the brackets mean the average over the phase space variables
$\r_2$, $\p_1$ and energy $\eps$ with a weight $\eps$, 
\be\l{psav}
\langle \frac{r_{\perp} ^2}{\varepsilon}\rangle =
\frac{\int {\rm d} \eps\; \eps\; 
\int {\rm d} \r_2 {\rm d} p_1 \;(r_{\perp} ^2/\varepsilon)\; 
f_{\rm scl}}{
\int {\rm d} \eps\; \eps \int {\rm d} \r_2 {\rm d} p_1\; f_{\rm scl}}.
\ee
Using now the same subdivision in terms of the ETF and shell components
for the MI (\ref{misclps}) and s.p.\ energy (\ref{escl}), after the 
statistical averaging for 
a finite temperature $T$ one finds 
\be\l{dmiF}
  \delta \Theta_{\rm scl} \approx  
           m \, \langle r_{\perp} ^2 /\varepsilon \rangle \delta F,
\ee
where $\delta F$ is the free-energy shell correction. Here we used 
the stationary phase (periodic orbit (PO)) conditions for the evaluation 
of integrals over the phase-space variables $\r_2$ and $\p_1$ \cite{MAF2006}.
Within the POT,
at a given temperature $T$, one has the PO sum 
\cite{strut,strutmag,KMS1979,FKMS1998,MSKB2010,belyaev}%
\be\l{dfscl}
 \delta F = \sum_{\rm PO} 
\frac{\pi t_{\rm PO} T / \hbar}
{\sinh (\pi t_{\rm PO} T / \hbar )}\;\delta E_{\rm PO},  
\ee
where $\delta E_{\rm PO}$ is the PO component of the 
energy shell-correction,
\bea\l{descl}
  \delta E &\approx& \delta E_{\rm scl} = \sum_{\rm PO} \delta E_{\rm PO}
  \qquad  \mbox{with}  \nonumber\\  
 \delta E_{\rm PO} &=& d_s \, \frac{\hbar ^2}{t_{\rm PO}^2} \; 
 \delta g^{}_{\rm PO}(\lambda).
\eea
Here 
$t^{}_{\rm PO}$ and $\delta g^{}_{\rm PO} (\eps)$ are 
the period of the particle motion along the PO (accounting for
its repetition number)  and the PO component
of the oscillating (shell) correction
to the level density, taken both 
at the chemical potential $\eps =\lambda$ for $\om=0$, 
which, at zero temperature, is equal
to the Fermi energy $\eps^{}_F$. This PO component
can be represented as
\be\l{dgscl}
\delta g_{\rm scl} (\eps) 
= \sum_{\rm PO} \delta g^{}_{\rm PO} (\eps),
\ee
where
\be\l{dgPO} 
 \delta g^{}_{\rm PO}(\eps)= \Re \left\{{\cal B}_{\rm PO} \;
\exp\left[\frac{i}{\hbar} S_{\rm PO}(\eps) -
\frac{i\pi}{2}\mu^{}_{\rm PO} \right]\right\},
\ee

\vspace{1.0cm}
\begin{figure}
\begin{center}
\includegraphics[width=0.47\textwidth]{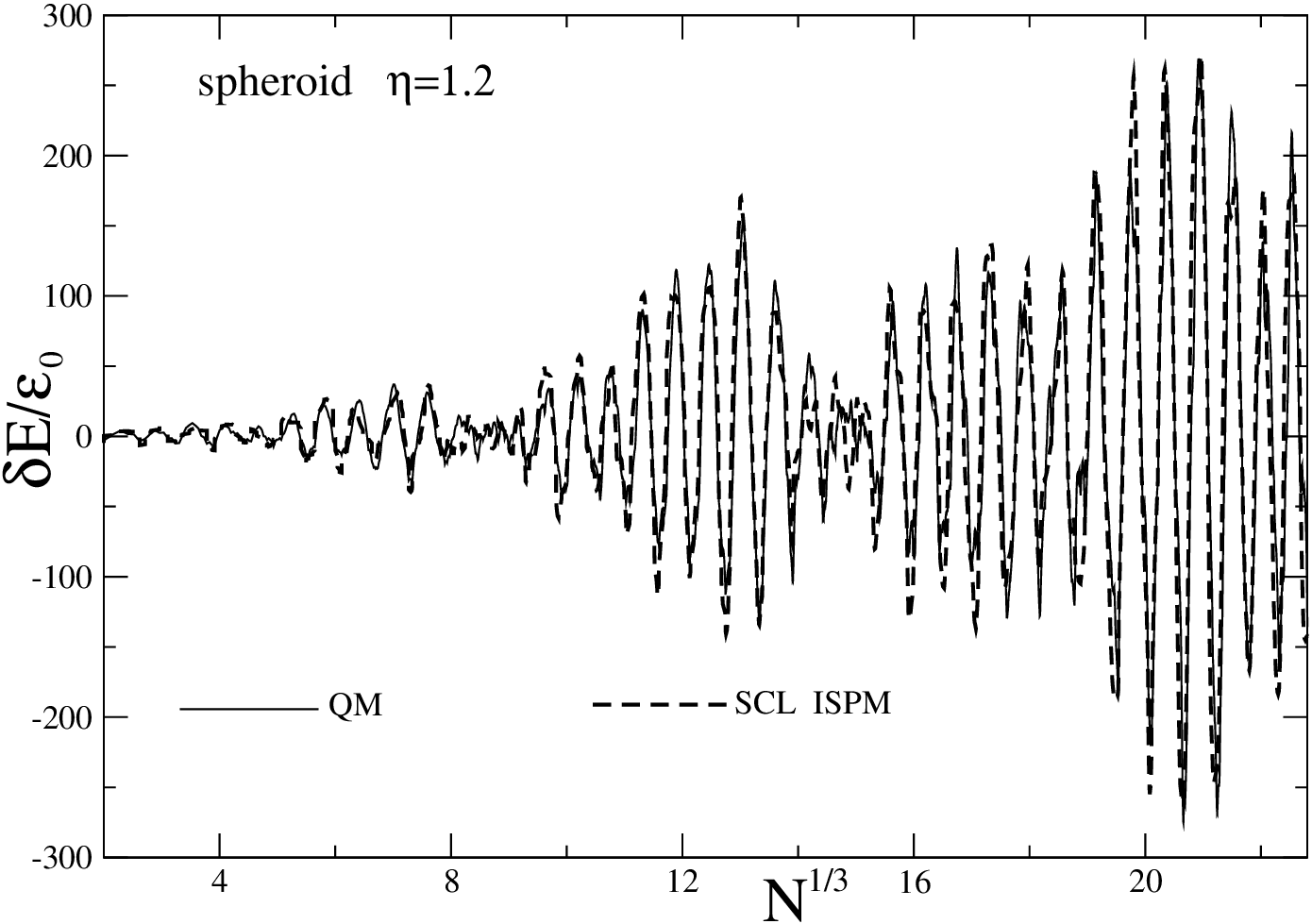}
\end{center}

\vspace{-0.6cm}
\caption{
Semiclassical (SCL ISPM, i.e. using the improved stationary phase method)
and
quantum (QM) shell corrections $\delta E$ versus the 
particle number variable $N^{1/3}$ for the spheroidal cavity potential 
(in units of $\eps_0=\hbar^2/2mR^2$) for the deformation 
$\eta=b/a=1.2$ (after \cite{MAFM2002}).
}
\label{fig2}
\end{figure}

\noindent 
with
${\cal B}_{\rm PO}$ the amplitude of the density oscillations
depending on the PO classical degeneracy and stability, and
 $S_{\rm PO}(\eps)$ the action along the PO 
\cite{gutz,strut,strutmag,book,belyaev,MYAB2011}. 
In (\ref{gscl}), $g^{}_{\rm ETF}(\eps)$ is the smooth ETF component and 
$ \delta g_{\rm scl}(\eps)$ the semiclassical oscillating contribution  
\cite{book} where the latter can be expressed in terms of the PO sum, 
Eqs.\  (\ref{dgscl}) and (\ref{dgPO}). The POs appear 
through the stationary phase 
(PO) condition for the calculation of the 
   integrals 
over $\r_2$ and $\p_1$ in (\ref{gscldef})
by the improved stationary phase method 
\cite{MAFM2002,MAF2006,MYAB2011} as mentioned above.
For the phase space average 
$\langle r_{\perp}^2 / \eps \rangle$, Eq.\ (\ref{psav}),
one again obtains approximately, through the Green's function (\ref{psgfun}), 
a decomposition
into an ETF and a shell-correction contributions:
\be\l{ell2}
  \langle r_{\perp}^2 / \eps \rangle 
  \approx \langle r_{\perp}^2 / \eps \rangle^{}_{\rm ETF} + 
  \delta \langle r_{\perp}^2 / \eps \rangle .
\ee

\section{Comparison with quantum calculations}

\begin{figure}
\begin{center}
\includegraphics[width=0.47\textwidth]{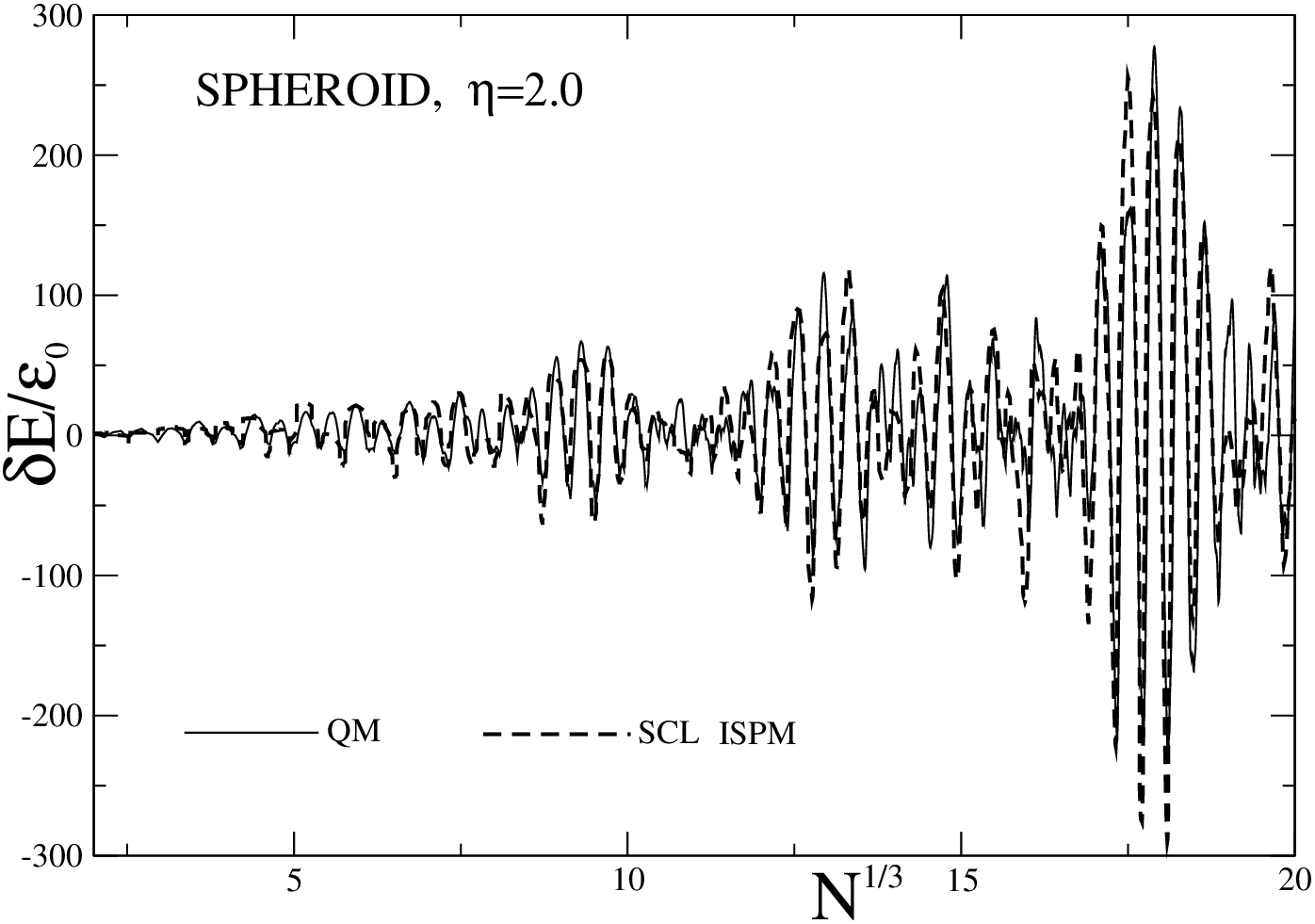}
\end{center}

\vspace{-0.4cm}
\caption{
The same as in Fig.\ \ref{fig2} but for a large deformation 
$\eta=2.0$ (after \cite{MAFM2002}).}
\label{fig3}
\end{figure}

In Figs.\ \ref{fig2} and \ref{fig3}, we show the quantum and semiclassical 
energy shell corrections $\delta E$ for the spheroidal cavity at small
(almost no contributions from PO bifurcations) and large deformations 
(important contributions of PO bifurcations). At zero temperature, the energy 
shell correction $\delta E$ corresponds to the free-energy shell correction 
$\delta F$. With increasing temperature $T$, one observes an 
exponential decrease of the free-energy shell
corrections $\delta F \sim \exp(-\pi t^{}_{\rm PO}T/\hbar)$ 
\cite{KMS1979,FKMS1998,MSKB2010,belyaev}. Minima of the energy shell corrections
$\delta E$ are related to magic particle numbers. 

Evaluating the 
factor $\langle r_{\perp}^2 / \eps \rangle$ of Eq.\ (\ref{psav})
in (\ref{dmiF}) 
within 
the simplest TF approximation for the level density

\begin{figure}
\begin{center}
\includegraphics[width=0.47\textwidth]{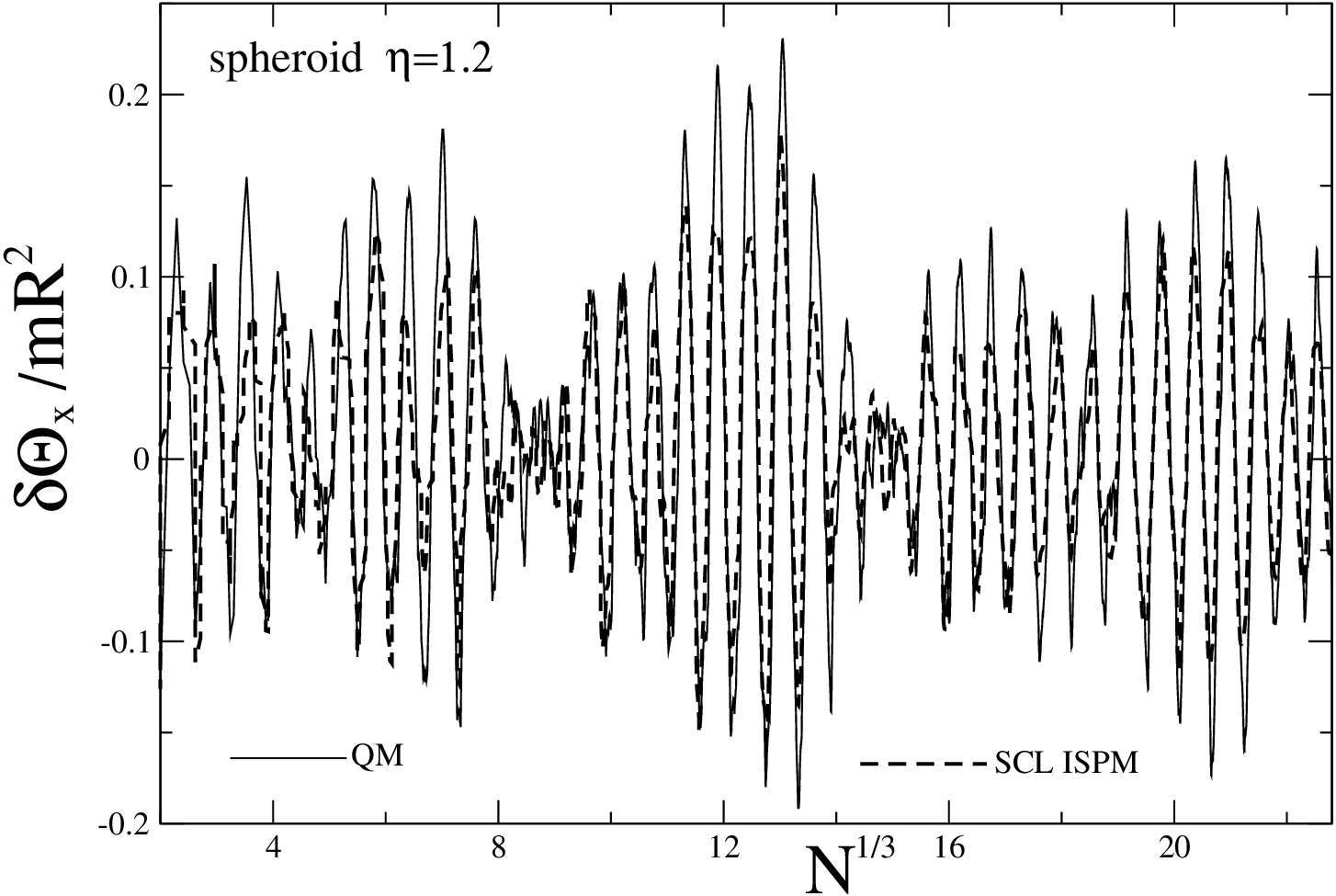}
\end{center}

\vspace{-0.4cm}
\caption{
Comparison of quantum (QM) and semiclassical (SCL ISPM) 
moment of inertia shell corrections 
$\delta \Theta_x$ (in single-particle units $m R^2$) in the case of the 
perpendicular collective rotation at zero temperature 
as function of $N^{1/3}$ at the same deformation as in Fig.\ \ref{fig2}.}
\vspace{ 0.8cm}
\label{fig4}
\end{figure}

\begin{figure}
\begin{center}
\includegraphics[width=0.47\textwidth]{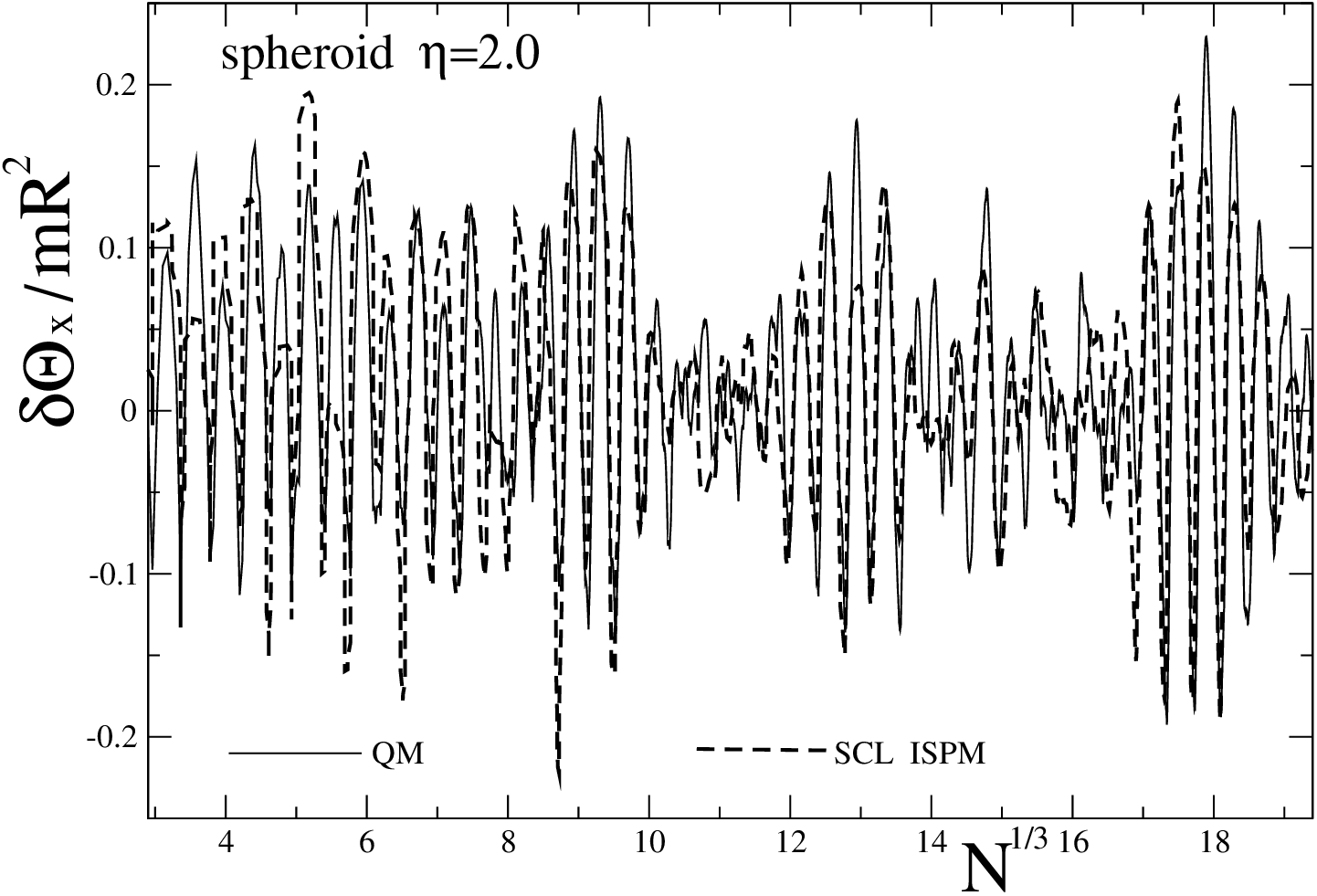}
\end{center}
\vspace{-0.0cm}
\caption{
Same as Fig.\ 4 but for a deformation $\eta=2.0$. } 
\label{fig5}
\end{figure}

$g_{\rm scl}(\eps)\approx g^{}_{\rm TF}(\eps) $ (\ref{gscl})
and the TF spectral function $f_{\rm scl} \approx f^{}_{\rm TF}$ 
(\ref{distrfuntf})
corresponding to the CT$_0$ component of the Green's function (\ref{psgfun})  
in the nearly local approximation,
one can simply neglect  $\hbar$
corrections (surface and curvature terms) and shell effects. 
For the spheroidal cavity potential, 
one obtains from (\ref{psav}) and (\ref{distrfuntf})
\be\l{avtf}
  \langle \frac{r_{\perp}^2}{\eps} \rangle 
    \approx \langle \frac{r_{\perp}^2}{\eps} \rangle_{\rm TF}
  = \frac{a^2 + b^2}{3 \lambda},
\ee
where $a$ and $b$ are the semi-axises of the spheroidal cavity,
$a^2b=R^3$, $R$ is the radius of the equivalent sphere.  
Using this estimate, one may evaluate 
the MI shell correction for two selected deformations $\eta=b/a$
(see Fig.\ \ref{fig4} and \ref{fig5}).                                                                  

 One notices that 
there is a large supershell effect in both $\delta E$ and 
$\delta \Theta_x$. The shell structure is much enhanced by
 the bifurcations
of the shorter 3-dimensional (3D) orbits from the parent equatorial (EQ) orbits
\cite{MAFM2002} at $\eta=2$.
Both kinds of PO families yield the essential 
contributions through $\delta F$, in contrast to the classical perturbation 
results of \cite{DFP2004} where the EQ orbits do not contribute 
and 3D PO contributions are not considered.
Our non-perturbation results for the MI shell corrections can be applied for
larger rotational frequencies and larger deformations 
($\eta \!\sim\! 1.5 - 2.0$) where the 
bifurcations play the do\-minating role like in the case of the deformed 
harmonic oscillator \cite{MSKB2010}.

\section{Conclusions}

We derived the shell components $\delta \Theta$ of the 
moment of inertia 
in terms of the free-energy shell correction $\delta F$ within the  
non-perturbative extended Gutzwiller POT for any effective mean-field 
potential using the phase-space variables. 
For the deformed spheroidal cavity and the harmonic oscillator potentials, 
we found a good agreement
between the semiclassical POT and quantum results for  
$\delta F$ and $\delta \Theta$ 
using the Thomas-Fermi approximation for 
$\langle r_{\perp}^2/\eps \rangle$ 
at several critical deformations and temperatures.
For smaller temperatures a remarkable interference of the 
dominant short three-dimensional and parent equatorial orbits and 
their bifurcations in the superdeformed region are shown.
For larger temperatures, the shorter EQ orbits are dominant in this 
comparison. An exponential decrease of the shell corrections with 
increasing temperature is analytically demonstrated.

\begin{ack}
We thank S.\ Aberg, K.\ Arita, R. K.\ Bhaduri, M.\ Brack,
F. A.\ Ivanyuk, S. N.\ Fedotkin, S.\ Frauendorf, M.\ Matsuo, K.\ Matsuyanagi, 
V. O.\ Nesterenko, V. V.\ Pashkevich, V. A.\ Plujko, K.\ Pomorski, 
and A. I.\ Sanzhur for many helpful and stimulating
discussions. One of us (A.G.M.) is
also very grateful for a nice hospitality and financial support 
during his working visits of the
National Centre for Nuclear Research in Otwock-Swierk and Warsaw,
the Institut Pluridisplinaire Hubert Curien in Strasbourg, the Nagoya
Institute of Technology, and
the Japanese Society of Promotion of Sciences, 
ID No. S-14130. 
\end{ack}

\end{document}